\documentclass[twocolumn,showpacs,preprintnumbers,amsmath,amssymb]{revtex4}

\usepackage{graphicx}
\usepackage{dcolumn}
\usepackage{bm}

\begin{document}


\title{Room-Temperature Ferromagnetism in Co-Doped TiO$_2$ Anatase: Role of Interstitial Co}

\author{W. T. Geng$^{1,2}$}
\author{Kwang S. Kim$^1$}%
\affiliation{%
$^1$National Creative Research Initiative Center for Superfunctional Materials and Department of Chemistry, Pohang University of Science and Technology, Pohang 790-784, Korea \\
$^2$Fritz-Haber-Institut der Max-Planck-Gesellschaft, Faradayweg 4-6, D-14195 Berlin, Germany
}%

\date{\today}

\begin{abstract}
 TiO$_2$ anatase doped with Co has been recently reported to exhibit  room-temperature ferromagnetism. 
$Ab$ $initio$ study on substitutional Co doping, however, yielded much larger magnetic moment for Co than experiment.
Our calculations based on density-functional theory show that the substitutional Co ions incorporated into TiO$_2$ anatase tend to cluster and then the neighboring interstitial tetrahedral sites become energetically favorable for Co to reside, yielding a local environment more like Co$_3$O$_4$ than CoTiO$_3$. 
The interstitial Co destroys the spin-polarization of the surrounding substitutional Co but enhances the stability of the ferromagnetism significantly. 
In the absence of carriers, this room-temperature ferromagnetism can only be accounted for by superexchange interaction.

\end{abstract}

\maketitle
The discovery of ferromagnetism in III-V semiconductors and the successful control of spin coherence of electrons injected from a magnetic semiconductor into a nonmagnetic semiconductor suggests the possibility of harnessing both charge and spin for new functionalities\cite{1,2,3}.
The limiting factor that represents a serious bottleneck for their practical spintronic applications is the fact that both the observed ferromagnetism and the attractive injection phenomena are essentially limited to low temperature.
The development of new materials grows rapidly. Room-temperature ferromagnetism has been observed, for instance, in Mn doped ternary compounds such as CdGeP$_2$\cite{CdGeP2}, ZnSnAs$_2$\cite{ZnSnAs2}, and ZnGeP$_2$\cite{ZnGeP2}. 

Using combinatorial pulsed-laser-deposition (PLD) molecular beam epitaxy (MBE) technique, Matsumoto $et$ $al$.\cite{science} have fabricated Co-doped anatase thin films with different Co contents on SrTiO$_3$(001) surface. At a concentration of 0.07, the film was reported to be ferromagnetic at room-temperature with a magnetic moment of 0.32 $\mu_B$ per Co.
In a more recent work, Chambers  $et$ $al$.\cite{apl} employed oxygen-plasma-assisted (OPA) MBE to grow Co-doped anatase films on the same substrate. They observed a significantly larger magnetic moment for Co ions, i.e., 1.26  $\mu_B$, which is much closer to the value of the low spin state of Co. From the comparison of the surface-sensitive X-ray photoemission spectrascopy (XPS) and bulk-sensitive X-ray absorption and emission spectroscopy (XAS) of Co-doped TiO$_2$ anatase and various other reference compounds such as Co, Co$_2$O$_3$, Co$_3$O$_4$, and CoTiO$_3$, Co ions in anatase were judged to be substitutional and in the +2 formal oxidation state. Also, they found the Co distribution, and therefore the magnetic properties depend critically on the growth condition.

To gain a microscopic understanding of the magnetism of this noval materials, Park  $et$ $al$.\cite{prb} carried out a systematic computational study on Ti$_{0.9325}$$M_{0.0625}$O$_2$ ($M$=Co, Mn, Fe, and Ni) using the linearized muffin-tin orbital (LMTO) method both with  local-spin-density approximation (LSDA) and with LSDA + $U$ incorporating the  on-site Coulomb correlation interaction $U$. Their calculations showed that the spin magnetic moment of Co was about 1$\mu_B$, indicating a low spin state. But they also obtained an orbital moment for Co as large as 0.9$\mu_B$, so the total magnetic moment per Co was 1.90$\mu_B$, about six times as high as the PLD-MBE result\cite{science}, or, 50$\%$ higher than the OPA-MBE experiment\cite{apl}.
This discrepancy is obviously too large to be acceptable. Recent  density-functional computations on various diluted magnetic semiconductors such as Mn-doped GaAs\cite{t1,t2} and ZnGeP$_2$\cite{ZnGeP2}, for instance, have reproduced a magnetic moment of Mn in good agreement with experiment\cite{3,ZnGeP2}. Although LSDA is not appropriate to describe strongly correlated systems, LSDA+$U$ has proved quite successful to yield a valid magnetic structure for such systems\cite{CoO}. 
Thus, the marked disagreement between experiment and theory casts some doubt on the understanding of the magnetism and its connection to the structure of Co-doped anatase. 
Motivated by the recent discovery that defects could play a key role in the magnetic structure of the diluted magnetic semiconductors\cite{zunger,eriksson}, we 
conjecture that clustering of substitutional Co and/or interstitial Co, if exists, might be responsible for the reduction of Co magnetization.

In an attempt to clarify this point, we have performed $ab$ $initio$ density-functional calculations to study the structural, electronic, and magnetic properties of Co-doped TiO$_2$ anatase. We show that the substitutional Co ions tend to cluster, but this has no remarkable effect on the magnetization of Co. We also find that a (seriously flattened) tetrahedral interstitial site will become energetically favorable for a Co atom (in reference to bulk Co) when two of its neighboring substitutional (octahedral) sites are occupied by Co. 
The interstitial Co is also in the low spin state (1.04$\mu_B$), but it exerts strong detrimental effect on the magnetization of its neighboring Co (0.69$\mu_B$ $\rightarrow$ $-0.04\mu_B$) and therefore reduces the magnetic moment Co in this magnetic semiconductor. 
While substitutional Co ions turn anatase half-metallic, the interstitial ones bring the material back to a semiconductor. In absence of carriers, the exchange interaction stablizing the ferrimagnetism is of superexchange type. 
In the following, we give a brief outline of our theoretical procedure, a discussion of the results based on bonding characteristics and its relevance to our understanding of the magnetism in this Co-doped oxide.

The calculations were done using a spin-polarized version of the Vienna {\it ab initio} Simulation Package (VASP)\cite{vasp1}. Fully nonlocal Vanderbilt-type ultrasoft pseudopotentials\cite{pp} were used to represent the electron-ion interaction. Exchange correlation interactions were described by the Perdew-Wang 1991 generalized gradient approximation (GGA)\cite{pw91}. 
Wave functions and  augmentation charges were expanded in terms of plane waves with an energy cutoff of 396 and 700 eV, respectively. The Brillouin zone integration was performed within Monkhorst-Pack scheme\cite{k} using a (2$\times$2$\times$2) $k$ mesh for geometry optimization and a (4$\times$4$\times$4) mesh for plotting the density of states(DOS).  
For each system, the geometry was relaxed until the atomic forces were smaller than 0.03 eV/\AA.

To gain a confidence for applying pseudopotentials to such an oxide with open structure, we first conducted a comparitive study of the structural properties of non-doped anatase (see Figure 1a) using both VASP and the highly precise all-electron full-potential linearized augmented plane wave (FLAPW) method\cite{flapw} with WIEN97 implementation\cite{wien}. FLAPW is considered to be more accurate but also more computationally expensive. The same GGA functional and $k$ mesh ($8\times 8\times 4$) were used in both calculations. The optimized structural parameters for anatase are listed in Table I. 
Also listed are experimental values\cite{tio2} and a previous FLAPW + LDA result\cite{asahi}. It is evident that VASP agrees fairly well with WIEN97, and that GGA results are more consistent with experiment than LDA. 

For Co-doped anatase, we mainly dealt with a supercell containing 16 primitive unit cells ($2a\times 2a\times c$, see Figure 1b) with one or two Ti atoms (site B1, or, sites B1 and B2) replaced by Co and with/without an interstitial Co (site A). In view of the fact that the calculated $c/a$ value, 2.46, of a hyperthetical CoO$_2$ in anatase structure is close to that of TiO$_2$ anatase, we fixed $c/a$ at the latter value when optimizing the volume of Co-doped anatase to save computational effort.
To see whether substitutional Co (Co$^S$) cluster or not, we first minimized the free energy of the cell Ti$_{15}$Co$^S_{B1}$O$_{32}$. Upon Co-doping, the volume of such a unit cell contracted by 1.1$\%$, in accordance with the fact that Co has a smaller covalent radius than Ti\cite{pauling}. Then we doubled this cell along $x$ direction and brought the two Co$^S$ together to form Ti$_{30}$Co$^S_{B1}$Co$^S_{B2}$O$_{64}$ ($4a\times 2a\times c$). We assumed that there was no volume change upon clustering, and optimized only the internal freedoms for the 96 atom cell. We find the latter configuration is more stable by 0.14eV/Co, indicating that Co clustering does occur.

The calculated formation energy of an interstitial Co (Co$^I$) as a neutral defect, $E$(Ti$_{16}$Co$^I_A$O$_{32}$)-$E$(Ti$_{16}$O$_{32}$)-$E$(Co), is +2.07 eV, implying a sole Co$^I$ is highly unstable in reference to bulk Co. This can be understood in the bond-order~band-strength picture. Ti:O + Co:O is much weaker than O:Ti:O, as is evident from the comparison of the formation heat of TiO$_2$, TiO, and CoO\cite{CRC}.
 Nonetheless, with the presence of Co$^S_{B1}$, the formation energy of Co$^I$, i.e., $E$(Ti$_{15}$Co$^S_{B1}$Co$^I_A$O$_{32}$)-$E$(Ti$_{15}$Co$^S_{B1}$O$_{32}$)-$E$(Co), drops drastically to +0.18 eV. According to our calculations, the formation heat of CoO$_2$ in a hypothetical anatase structure is $-3.66$ eV, slightly higher than two times of that of CoO ($-1.59$ eV). This means that  Co:O + Co:O is only slightly weaker than O:Co:O , thus Co$^I$ becomes much less unstable.  Further calculations show that when both B1 and B2 site are occupied by Co upon clustering, site A becomes favorable for a Co$^I$ with a formation energy of $-0.49$ eV. Once again, this can be explained by the higher formation heat of Co$_3$O$_4$ ($-9.57$ eV) than two times of that of CoO$_2$ ($-3.66$ eV).
Our calculations thus reveal a 2Co$^S$+Co$^I$ local structure in Co-doped TiO$_2$ anatase.

Table II displays the calculated Co-O bond length in different local environments. The overall feature is that the Co-O bond is shorter than Ti-O bond. As mentioned above, this is because both octahedral and tetrahedral Co ions have a radius smaller than octahedral Ti ions. 
Co-O bond lengths along the $y+$ and $y-$ directions, $d_{y+}$ and $d_{y-}$, do not change upon clustering in the case that Co(B1) and Co(B2) are in the same $x-z$ plane; whereas along the $x$ and $z$ directions the bond lengths are contracted by $0.02-0.05$ \AA. The presence of an interstitial Co pushes the surrounding atoms a bit further apart and increases the Co-O bond length in the case of Co(B1) and Co(B2). The local lattice distortion induced by Co$^S$ or Co$^I$, however, is rather small. It is worth noting that the volume effect on chemical bonding is only marginal. A sole Co$^I$ yields a volume expansion of about 1.0$\%$ to Ti$_{16}$O$_{32}$, and the formation heat of Ti$_{16}$Co$^I_A$O$_{32}$ at the volume of Ti$_{16}$O$_{32}$ is only 0.05 eV lower than the optimized value.

To examine the effect of clustering on magnetism, we performed calculations on both intra-cell FM and intra-cell AFM alignments of the Co pair in a Ti$_{14}$Co$^S_{B1}$Co$^S_{B2}$O$_{32}$ unit cell (For simplicity, we denote this cell as Ti$_{14}$Co$_2$O$_{32}$ in the following) while keeping the inter-cell coupling ferromagnetic. 
The FM phase is found to be lower in total energy by 0.09 eV than the AFM phase. When the two Co ions are far apart, this energy difference is 0.08 eV\cite{prb}. Apparently, clustering has negligible effect on the spin-polarization of this magnetic oxide. In the case of 2Co$^S$+Co$^I$, we checked all the three possible intra-cell spin alignments of the three Co ions. All of them converge to the same final state, with both Co$^I$ and Co$^S$ in the low spin state. We then double the unit cell along $x$ direction and computed the inter-cell AFM coupling between two  2Co$^S$+Co$^I$ complexes. It is found to be 0.35 eV higher in energy than the FM coupling. Obviously, with a much enhanced FM-AFM energy difference, the formation of  2Co$^S$+Co$^I$ will raise remarkably the Curie temperature of Co-doped anatase, in accordance with the observed room-temperature ferromagnetism\cite{science,apl}.

The calculated total spin magnetic moment in a unit cell of Ti$_{15}$Co$_1$O$_{32}$, Ti$_{14}$Co$_2$O$_{32}$, and Ti$_{14}$Co$_3$O$_{32}$, and that in the atomic sphere of Co are listed in Table III. 
It is seen that in all three cases, Co ions are in the low spin state, regardless to their oxidation states. The spin moment of Co$^S$ does not change much upon clustering. Co$^I$, on the other hand, behaves rather differently. When coupled with Co$^S$, Co$^I$ kills almost entirely the spin moments of  Co$^S$. As a result, the average spin magnetic moment of Co is reduced to 0.32 $\mu_B$, less than a half of the non-Co$^I$ case. This means that if there are a remarkable amount of  2Co$^S$+Co$^I$ complexes in Co-doped anatase, the magnetic moment per Co atom would be reduced significantly and this could explain the disagreement between experiments\cite{science,apl} and the previous first-principles calculations\cite{prb}.
We note that the oxygen-rich condition during OPA-MBE growth in Chambers $et$ $al.$'s work\cite{apl} prevented the formation of extensive  2Co$^S$+Co$^I$ complexes, hence a much higher magnetic moment for Co than that in the film grown by Matsumoto $et$ $al.$\cite{science} with PLD-MBE; first-principles calculations\cite{prb} gave a even higher magnetic moment for Co$^S$, as a result of the absence of  Co$^I$.
In oxygen-poor condition, on the other hand, the oxygen vacancies help the diffusion of the Co ions, resulting in the formation of Co metal clusters\cite{cluster}.

Figure 2 displays the calculated density of states for non-doped and various cases of Co-doped anatase. Clearly, the Co states are mainly located in the energy-gap region and the O and Ti states are not much affected by Co doping. A comparison of Ti$_{15}$Co$_1$O$_{32}$ and Ti$_{14}$Co$_2$O$_{32}$ indicates clustering does not change the half-metallic electronic structure, and as a consequence the magnetic interaction between Co ions remains to be carrier-mediated Zener-like.\cite{zener} Sole Co$^I$ ions turn the material into a metallic system with a highly lifted fermi level. Very interestingly, our calculations show a complex of  2Co$^S$+Co$^I$ brings the system back to a semiconductor, in consistence with the experimental observation\cite{science,apl}.  

To gain a deeper insight into the electronic structure of Co, we compare $d$ DOS of Co in anatase with those of Co in other Co oxides with different oxidation states such as CoO$_2$ (+4), CoTiO$_3$ (+2), and Co$_3$O$_4$ (+3 and +2). CoO$_2$, the end member of LiCoO$_2$, has recently been isolated\cite{coo2_1}. It seems to be of CdCl$_2$ type with a monoclinic distortion, but positions of O remain to be determined. In this regard, we believe a study on a hypothetical CoO$_2$ in anatase structure is meaningful in illustrating the electronic structure of Co$^{4+}$. As mentioned above, we have optimized all of its structural parameters.
For CoTiO$_3$ and Co$_3$O$_4$, experimental lattice constants\cite{wyckoff} were adopted and only internal freedoms were optimized. The calculated $d$ DOS of Co in  CoO$_2$, CoTiO$_3$, and Co$_3$O$_4$, and that in Co-doped anatase are plotted in Figure 3. We observe that although much more localized, $d$ DOS of Co$^S$ in Ti$_{15}$Co$_1$O$_{32}$ and Ti$_{14}$Co$_2$O$_{32}$ resembles that of Co in  CoO$_2$ in both exchange splitting and crystal field splitting, suggesting they are in +4 oxidation state, as was also proposed in Ref.[9]. In an octahedral field, $t_{2g}$ is lower than $e_g$ and thus Co$^{4+}$ will have a spin magnetic moment of 1$\mu_B$. 
On the other hand, Co$^I$ in  Ti$_{16}$Co$_1$O$_{32}$ reminds us of Co in  CoTiO$_3$. This resemblance reveals  Co$^I$ is in +2 oxidation state in a  Ti$_{16}$O$_{32}$ environment. In a normal tetrahedral field, $e_g$ level is lower than  $t_{2g}$. Co$^I$, nevertheless, is in a seriously flattened tetrahedral field, thus the crystal field splitting states is comparable to the exchange splitting, yielding an intermediate spin state for Co$^I$. As for the combination  2Co$^S$+Co$^I$, an even better similarity can be found between Co$^S$ (Co$^I$) in anatase and Co$^B$ (Co$^A$) in Co$_3$O$_4$. It thus provides a strong evidence that  Co$^S$ and Co$^I$ in such a combination are in +3 and +2 oxidation states, respectively. Similar to  Co$^B$ in Co$_3$O$_4$,  Co$^S$ in an octahedral crystal field has a vanishing spin-polarization, and the contribution to magnetism comes only from  Co$^I$.  

To conclude, our prediction of the formation of 2Co$^S$+Co$^I$ resolves the disagreement between experiment and theory on the magnetic moment of Co in Co-doped TiO$^2$ anatase. The insulating property is reproduced without employing an on-site Coulomb correlation interaction $U$. 
Superexchange, rather than carrier-mediated Zener-like magnetic interaction, should be responsible for room-temperature ferromagnetism. 
The novel strucure discoverd in this work may also exist in other transition elements such as Mn- and Fe-doped anatase and may thus resolve the discrepancy between experiment\cite{ass} and theory\cite{prb}. Finally, we want to note that the formation of more complicated (Co$^S$, Co$^I$) combinations is also possible depending on the growth condition, but the formation probability should be lower than  2Co$^S$+Co$^I$. 
     
\begin{acknowledgments}
Work in Korea was supported by Korea Institute of Science and Technology Evaluation and Planning (Creative Research Initiative) and in Germany by Max-Planck-Society Fellowship. W.T.G. is grateful to Professor B. I. Min for helpful discussions.
\end{acknowledgments}

 \begin{figure}
 \caption{\label{fig1} Calculational unit cells of non-doped (panel a) and Co-doped TiO$_2$ anatase (panel b). Large and small circles represent O and Ti (Co), respectively.}
 \end{figure}

 \begin{figure}
 \caption{Total density of states for non-doped and various cases of Co-doped TiO$_2$ anatase. Dotted vertical lines denote the Fermi energy.}
 \end{figure}

 \begin{figure}
 \caption{The calculated density of $d$ states of Co in CoO$_2$, CoTiO$_3$, Co$_3$O$_4$, and various cases of Co-doped TiO$_2$ anatase. Dotted vertical lines denote the Fermi energy.\label{fig3}}
 \end{figure}

\begin{table}
\caption{\label{tab:table1}
Optimized structural parameters for anatase TiO$_2$: A comparison of all-electron full-potential method (WIEN97), pseudopotential method (VASP), experiment\cite{tio2}, and a previous FLAPW LDA study\cite{asahi}.  }
\begin{ruledtabular}
\begin{tabular}{lcccc}
        &WIEN97  &VASP &Expt. & FLAPW  \\
        &(GGA)  &(GGA) & & (LDA) \\
\hline
$a$ (\AA) & 3.826  & 3.825 & 3.782 & 3.692 \\ 
$c$ (\AA) & 9.706  & 9.678 & 9.502 & 9.471 \\
$c/a$     & 2.537  & 2.530 & 2.512 & 2.566 \\
$u$       & 0.207  & 0.207 & 0.207 & 0.206 \\
\end{tabular}
\end{ruledtabular}
\end{table}

\begin{table}
\caption{\label{tab:table2}
Optimized Co-O bond lengths (\AA) in Co doped anatase TiO$_2$. It is denoted as $d_{x+}$ ($d_{y+}$, $d_{z+}$) when O is in the $x+$ direction of Co, and as $d_{x-}$ ($d_{y-}$, $d_{z-}$) when O is in the $x-$ direction of Co. Co(B1), Co(B2), and Co(A) denote Co ions occupying sites B1, B2, and A (see Figure 1b). }
\begin{ruledtabular}
\begin{tabular}{ccccccc}
\multicolumn{1}{c}{$d_{\rm Co-O}$ }& \multicolumn{1}{c}{Ti$_{15}$Co$_1$O$_{32}$} & \multicolumn{2}{c}{Ti$_{14}$Co$_2$O$_{ 32}$} &  \multicolumn{3}{c}{Ti$_{14}$Co$_3$O$_{ 32}$} \\ \cline {2-2} \cline{3-4} \cline {5-7}
        &Co(B1) &Co(B1) &Co(B2) &Co(B1) &Co(B2) & Co(A)  \\
\hline
$d_{x+}$ & 1.91 & 1.85 & 1.93 & 1.93 & 1.98 & 1.92  \\
$d_{x-}$ & 1.91 & 1.93 & 1.85 & 1.88 & 1.93 & 1.91  \\
$d_{y+}$ & 1.91 & 1.91 & 1.91 & 1.96 & 1.97 & 1.91  \\
$d_{y-}$ & 1.91 & 1.91 & 1.91 & 1.90 & 1.93 & 1.85  \\
$d_{z+}$ & 1.89 & 1.84 & 1.95 & 1.93 & 1.98 &  -    \\
$d_{z-}$ & 1.89 & 1.94 & 1.84 & 1.96 & 1.98 &  -    \\
\end{tabular}
\end{ruledtabular}
\end{table}

\begin{table}
\caption{\label{tab:table3}
Claculated spin magnetic moment ($\mu_B$) in the atomic sphere of Co (radius = 1.10 \AA) and that for a whole unit cell. }
\begin{ruledtabular}
\begin{tabular}{ccccc}
Unit cell & Co(B1) & Co(B2) & Co (A) & Total \\ \hline
Ti$_{15}$Co$_1$O$_{ 32}$ & 0.73 & -   & -   & 1.00  \\
Ti$_{14}$Co$_2$O$_{ 32}$ & 0.69 & 0.69 & -  & 2.00  \\
Ti$_{14}$Co$_3$O$_{ 32}$ &-0.01 &-0.07 & 1.04 & 1.00 \\
\end{tabular}
\end{ruledtabular}
\end{table}

\bibliography{ti}

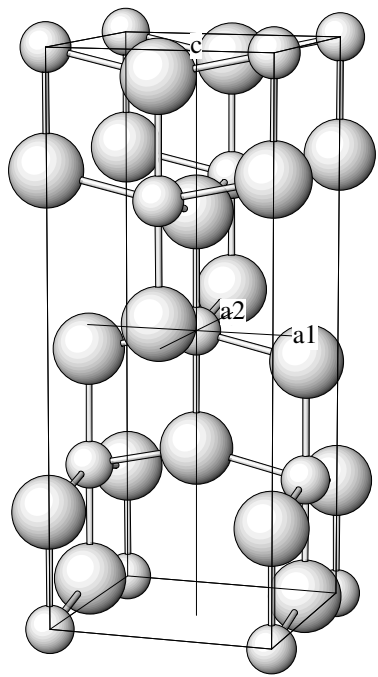
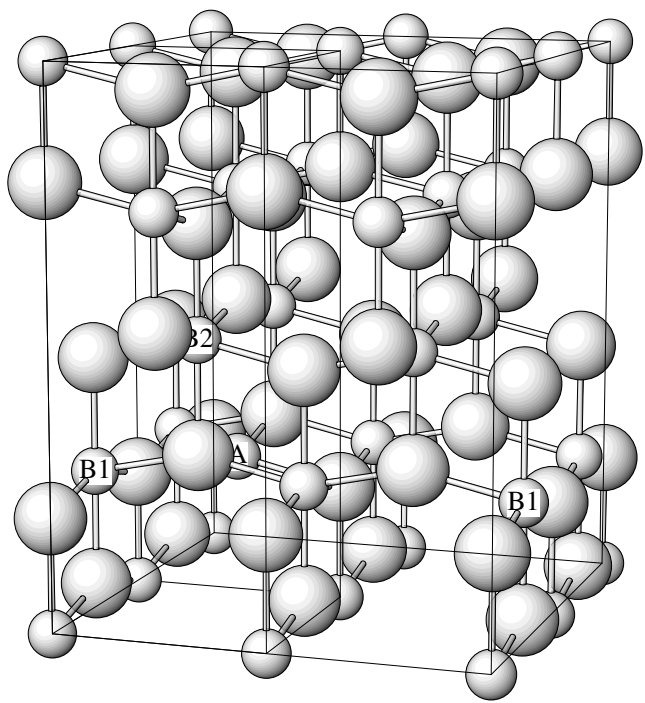
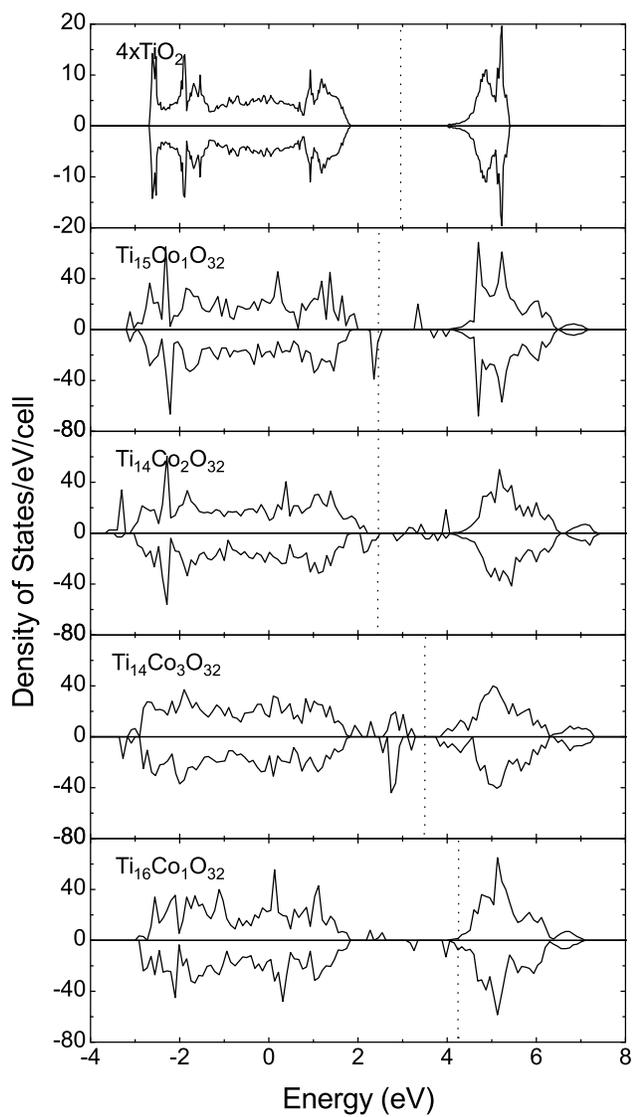
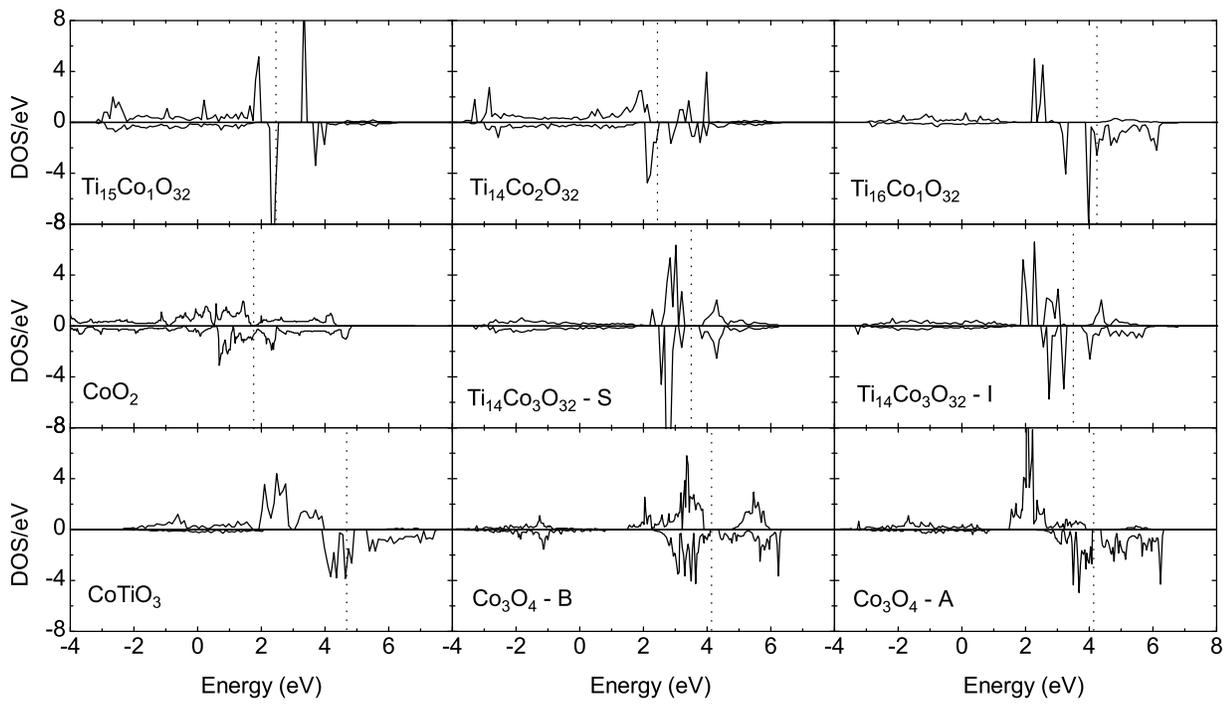
\end{document}